\pdfoutput=1
\documentclass[final,authoryear,5p,times]{elsarticle}

\usepackage{graphicx}
\usepackage[colorlinks]{hyperref}
\usepackage{amssymb}

\usepackage{listings}
\journal{Astronomy and Computing}

\begin{document}

\begin{frontmatter}



\title{Web-Based Visualization of Very Large Scientific Astronomy Imagery}


\author[upmc,cnrs]{Emmanuel Bertin}
\ead{bertin@iap.fr}
\author[c2rmf]{Ruven Pillay\corref{corr1}}
\ead{ruven.pillay@culture.gouv.fr}
\author[psud,cnrs]{Chiara Marmo}
\ead{chiara.marmo@u-psud.fr}

\address[upmc]{Univ Pierre et Marie Curie, Institut d'Astrophysique de Paris, UMR7095, Paris, F-75014, France}
\address[c2rmf]{C2RMF, Palais du Louvre - Porte des Lions, Paris 75001, France}
\address[psud]{Univ Paris-Sud, Laboratoire GEOPS, UMR8148, Orsay, F-91405, France}
\address[cnrs]{CNRS, France}

\cortext[corr1]{Corresponding author}

\begin{abstract}
Visualizing and navigating through large astronomy images from a remote location with current astronomy display tools can be a frustrating experience in terms of speed and ergonomics, especially on mobile devices.
In this paper, we present a high performance, versatile and robust client-server system for remote visualization and analysis of extremely large scientific images. Applications of this work include survey image quality control, interactive data query and exploration, citizen science, as well as public outreach. The proposed software is entirely open source and is designed to be generic and applicable to a variety of datasets. It provides access to floating point data at terabyte scales, with the ability to precisely adjust image settings in real-time. The proposed clients are light-weight, platform-independent web applications built on standard HTML5 web technologies and compatible with both touch and mouse-based devices. We put the system to the test and assess the performance of the system and show that a single server can comfortably handle more than a hundred simultaneous users accessing full precision 32 bit astronomy data.
\end{abstract}

\begin{keyword}
visualization \sep scientific data \sep web application \sep high resolution \sep HTML5 


\end{keyword}

\end{frontmatter}


\section{Introduction}
\label{chap:intro}

Although much of the extraction of information from astronomy science images is now performed ``blindly'' using computer programs, astronomers still rely on visual examination for a number of tasks. Such tasks include image quality control, assessment of morphological features, and debugging of measurement algorithms. 

The generalization of standardized file formats in the astronomy community, such as FITS \citep{wells_fits_1981}, has facilitated the development of universal visualization tools. In particular, {\sc SAOimage} \citep{1990BAAS...22..935V}, {\sc Aladin} \citep{1994ASPC...61..215B}, {\sc SkyCat} \citep{1997ASPC..125..333A}, {\sc Gaia} \citep{2000ASPC..216..615D} and {\sc ds9} \citep{2003ASPC..295..489J}. These packages are designed to operate on locally stored data and provide efficient access to remote image databases by downloading sections of FITS data which are subsequently read and processed locally for display; all the workload, including image scaling, dynamic range compression, color compositing and gamma correction, is carried out client-side.

However, the increasing gap between storage capacities and data access bandwidth \citep{budman} makes it increasingly efficient to offload part of the image processing and data manipulations to the server, and to transmit some form of pre-processed data to clients over the network. 

Thanks to the development of wireless networks and light mobile computing (tablet computers, smartphones), more and more scientific activities are now being carried out on-the-go outside an office environment. These possibilities are exploited by an increasing number of scientists, especially experimentalists involved in large international collaborations and who must interact remotely, often in real-time, with colleagues and data located in different parts of the world and in different time zones. Mobile devices have increasingly improved display and interfacing capabilities, however, they offer limited I/O performance and storage capacity, as well as poor battery life when under load. Web-based clients, or simply {\em Web Apps}, are the applications of choice for these devices, and their popularity has exploded over the past few years.

Thanks to the ubiquity of web browsers on both desktop and mobile platforms, {\it Web Apps} have become an attractive solution for implementing visual interfaces. Modern web browsers feature ever faster and more efficient JavaScript engines, support for advanced standards such as HTML5 \citep{html5} and CSS3 \citep{css}, not to mention interactive 3D-graphics with the recent WebGL API \citep{webgl}. As far as data visualization is concerned, web applications can now be made sufficiently feature-rich so as to be able to match many of the functions of standalone desktop applications, with the additional benefit of having instant access to the latest data and being embeddable within web sites or data portals.

One of the difficulties in having the browser deal with science data is that browser engines are designed to display gamma-encoded images in the GIF, JPEG or PNG format, with 8-bits per Red/Green/Blue component, whereas scientific images typically require linearly quantized 16-bit or floating point values. One possibility is to convert the original science data within the browser using JavaScript, either directly from FITS \citep{jsfits, astrojs}, or from a more ``browser-friendly'' format, such as e.g., a special PNG ``representation file'' \citep{js9}, or compressed JSON \citep{2011ASPC..442..467F}. In practice this is currently limited to small rasters, as managing millions of such pixels in JavaScript is still too burdensome for less powerful devices. Moreover, lossless compression of scientific images is generally not very efficient, especially for noisy floating-point data \citep[e.g.][]{2009PASP..121..414P}. Hence, currently, server-side compression and encoding of the original data to a browser-friendly format remains necessary in order to achieve a satisfactory user experience on the web client, especially with high resolution screens.

Displaying images larger than a few megapixels on monitors or device screens requires panning and/or pixel rebinning, such as in ``slippy map'' implementations (Google Maps\texttrademark, OpenStreetMap\footnote{\url{http://www.openstreetmap.org}} etc.). On the server, the images are first decomposed into many small tiles (typically $256\times 256$ pixels) and saved as PNG or JPEG files at various levels of rebinning, to form a ``tiled pyramid''. Each of these small files corresponds to a URL and can be loaded on demand by the web client. Notable examples of professional astronomy web apps based on this concept include the Aladin Lite API \citep{2012ASPC..461..443S}, and the Mizar plugin\footnote{\url{https://github.com/TPZF/RTWeb3D}} in SITools2 \citep{2012ASPC..461..821M}.

However, having the data stored as static 8-bit compressed images means that interaction with the pixels is essentially limited to passive visualization, with little latitude for image adjustment or interactive analysis. Server-side dynamic processing/conversion of science-data on the server and streaming of the processed data to the web client are necessary to alleviate these limitations. Visualization projects featuring dynamic image conversion/streaming in Astronomy or Planetary Science have mostly relied on browser plugins implementing proprietary technologies \citep{2012ASPC..461...95F} or Java clients/applets \citep{10.1109/MCSE.2009.142, kitaeff_2012}. Notable exceptions include Helioviewer
 \citep{hughitt_helioviewer:_2008}, which queries compressed PNG tiles directly from the browser with the tiles generated on-the-fly server-side from JPEG2000 encoded data.

In this paper we describe an open source and multi-platform high performance client-server system for the processing, streaming and visualization of full bit depth scientific imagery at the terabyte scale. The system consists of a light-weight C++ server and W3C standards-based JavaScript clients capable of running on stock browsers. In section \ref{chap:method}, we present our approach, the protocols and the implementation of both the server and the client. Sections \ref{chap:astrapp} and \ref{chap:planetapp} showcase several applications in Astronomy and Planetary Science. In Section \ref{chap:perf}, we assess the performance of the system with various configurations and load patterns. Finally in Section \ref{chap:conclu}, we discuss future directions in the light of current technological trends.

\section{Material and Methods}
\label{chap:method}

The proposed system consists of a (or several) central image server(s) capable of processing 32 bit floating point data on-demand and of transcoding the result into an efficient form usable by both light-weight mobile devices or desktop computers.

\subsection{Image Server}
\label{chap:server}

At the heart of the system is the open source {\sc IIPImage}\footnote{\url{http://iipimage.sourceforge.net}} image server \citep{PPL06}. {\sc IIPImage} is a scalable client-server system for web-based streamed viewing and zooming of ultra high-resolution raster images. It is designed to be fast, scalable and bandwidth-efficient with low processor and memory requirements.

\begin{figure}[h]
   \centering
   \includegraphics[width=\columnwidth]{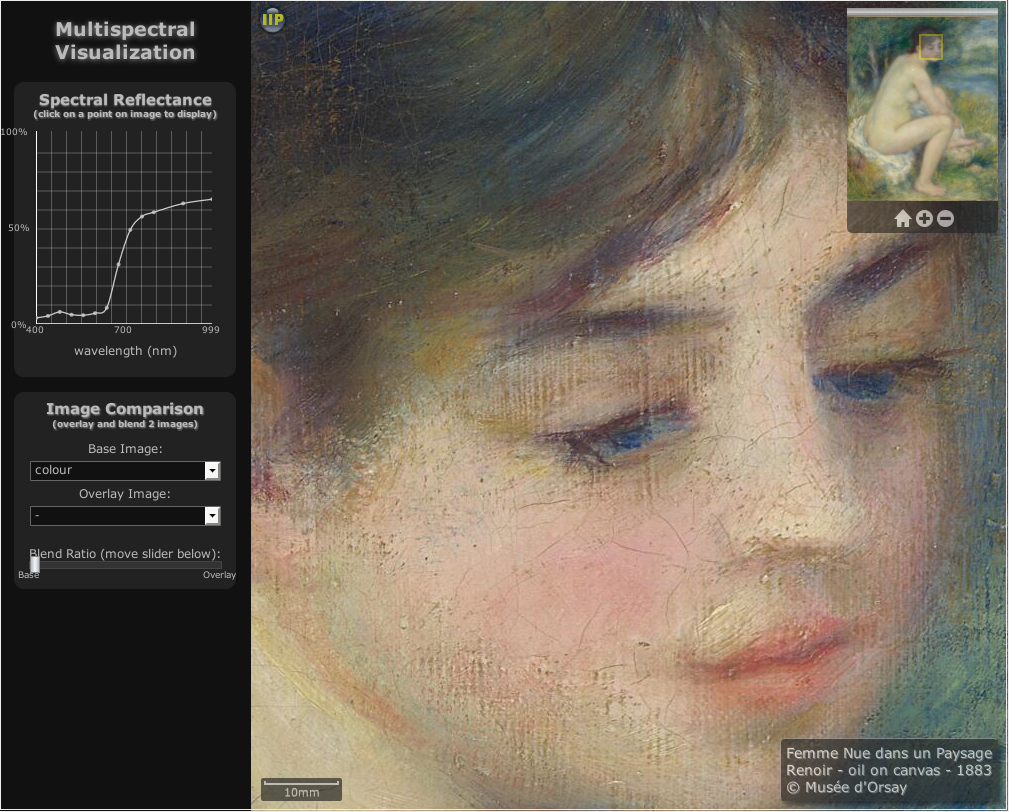}
      \caption{Spectral visualization of Renoir's \textit{Femme Nue dans un Paysage}, Mus\'{e}e de l'Orangerie, showing spectral reflectance curve for any location and controls for comparing different imaging modalities}
         \label{fig:multispectral}
   \end{figure}

\begin{figure}[h]
   \centering
   \includegraphics[width=0.9\columnwidth]{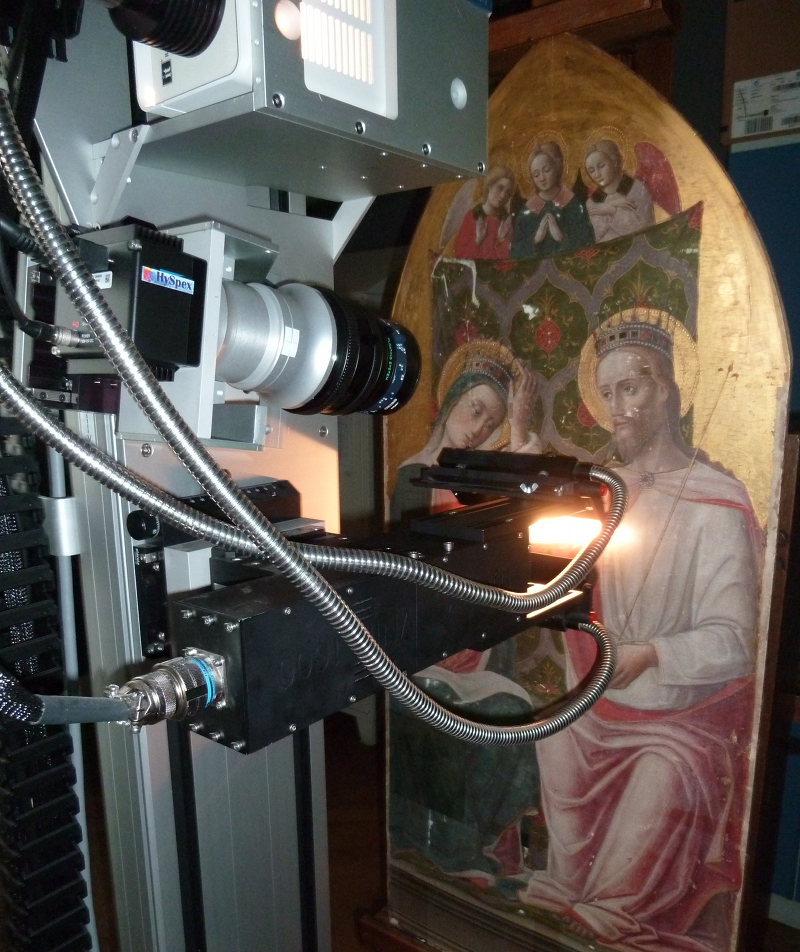}
         \caption{Hyperspectral imaging of paintings}
         \label{fig:hyperspectral}
   \end{figure}

{\sc IIPImage} has a long history and finds its roots in the mid 1990s in the cultural heritage field where it was originally created to enable the visualization of high resolution colorimetric images of paintings \citep{martinez_high_1998}. The original system was designed to be capable of handling gigapixel size, scientific-grade imaging of up to 16 bits per channel, colorimetric images encoded in the CIEL*a*b* color space and high resolution multispectral images \citep{martinez_ten_2002} (Fig. \ref{fig:multispectral}). It had hitherto been very difficult to simply even view such image data locally, let alone access it remotely, share or collaborate between institutions. The client-server solution also enabled integration of full resolution scientific imaging such as infra-red reflectography, Xray, multispectral and hyperspectral imagery (Fig. \ref{fig:hyperspectral}) into museum research databases, providing for unprecedented levels of interactivity and access to these resources \citep{lahanier_eros}.


Beyond cultural heritage, the system has also been adapted for use in the field of biomedical imaging. For example, to visualize ultra-large high resolution electron microscopy maps created by ultra-structural mapping or \textit{virtual nanoscopy} \citep{faas_virtual_2012}, or to explore high resolution volumetric 3D cross-sectional atlases \citep{husz_web_2012}.

In practice the {\sc IIPImage} platform consists of a light-weight C++ Fast-CGI \citep{fastcgi} server, {\tt iipsrv}, and an AJAX-based web interface. Image data stored on disk are structured in order to enable efficient and rapid multi-resolution random access to parts of the image, allowing terapixel scale data to be efficiently streamed to the client. As only the region being viewed needs to be decoded and sent, large and complex images can be managed without onerous hardware, memory or network requirements by the client. The {\sc IIPImage} server performs on-the-fly JPEG compression for final visualization, but as the underlying data is full bit depth uncompressed data, it can operate directly on scientific images, and perform operations such as rescaling or filtering before sending out the results to the client.

IIPImage, therefore, possessed many of the attributes necessary for astronomy data visualization and rather than develop something from scratch, it was decided to leverage this existing system and extend it. A further benefit to this approach would be access to a larger scientific community beyond that of astronomy with certain similar data needs. Moreover, as IIPImage forms part of the standard Debian, Ubuntu and Fedora Linux distributions, access to the software, installation and maintenance would be greatly simplified and sustainable in the longer term.

Hence a number of modifications were made to the core of {\sc IIPImage} in order to handle astronomy data. In particular, to extend the system to handle 32 bit data (both integer and IEEE floating point), FITS metadata, functionality such as dynamic gamma correction, colormaps, intensity cuts, and to be capable of extracting both horizontal and vertical data profiles. The resulting code has been integrated into the main {\sc IIPImage} software development repositories and is available from the project website\footnote{\url{http://iipimage.sourceforge.net}}, where it will form part of the $1.0$ release of {\tt iipsrv}.

\subsection{Data Structures and Format}
\label{chap:ipp}Extracting random image tiles from a very large image requires an efficient storage mechanism. In addition to tile-based access, the possibility to rapidly zoom in and out imposes some sort of multi-resolution structure on the data provided to the server. The solution adopted for ``slippy map'' applications is often simply to store individual tiles rebinned at the various resolution levels as individual image files. For a very large image this can translate into hundreds of thousands of small files being created. This approach is not convenient from a data management point of view, and for {\sc IIPImage} a single file approach has always been preferred.

The current version of {\sc IIPImage} supports both TIFF and JPEG2000 formats. Multi-resolution encoding is one of the major features of JPEG2000, but the lack of a robust, high performance open source library has been a serious issue until recently. Nevertheless, the encoding of floating point values spanning a large dynamic range remains a concern with current open-source libraries, as in practice input data is managed with only fixed point precision \citep{6092314,2014arXiv1403.2801K}.

The combining of tiling and multi-resolution mechanisms is also possible with TIFF. TIFF is able to store not only 8 bit and 16 bit data, but also 32 bit integers, and single or double precision floating point numbers in IEEE format. As a well supported and mature standard with robust and widely used open source development libraries readily available, TIFF was adopted as the main server-side storage format, rather than creating a completely new format or adapting existing science formats in some way.

\subsection{Image Transcoding}

Astronomy imaging data are usually stored in the FITS format \citep{wells_fits_1981}. FITS is a flexible container format that can handle data encoded in up to 64 bits per value. FITS supports image tiling, whereby the original raster is split into separate rectangular tiles, which can be retrieved quickly and read and decoded independently \citep{pence_w._fits_2000}. Versions of the same image could be stored at multiple resolution levels in different extensions, at the price of an increased file size. However currently neither tiling nor multi-resolution is present in archived FITS science images. Hence, regardless of the adopted storage format (TIFF in our case), a considerable amount of pixel shuffling and rebinning must be carried out in order to convert FITS data before they can be handled by the server.

\begin{figure}
   \centering
   \includegraphics[width=\columnwidth]{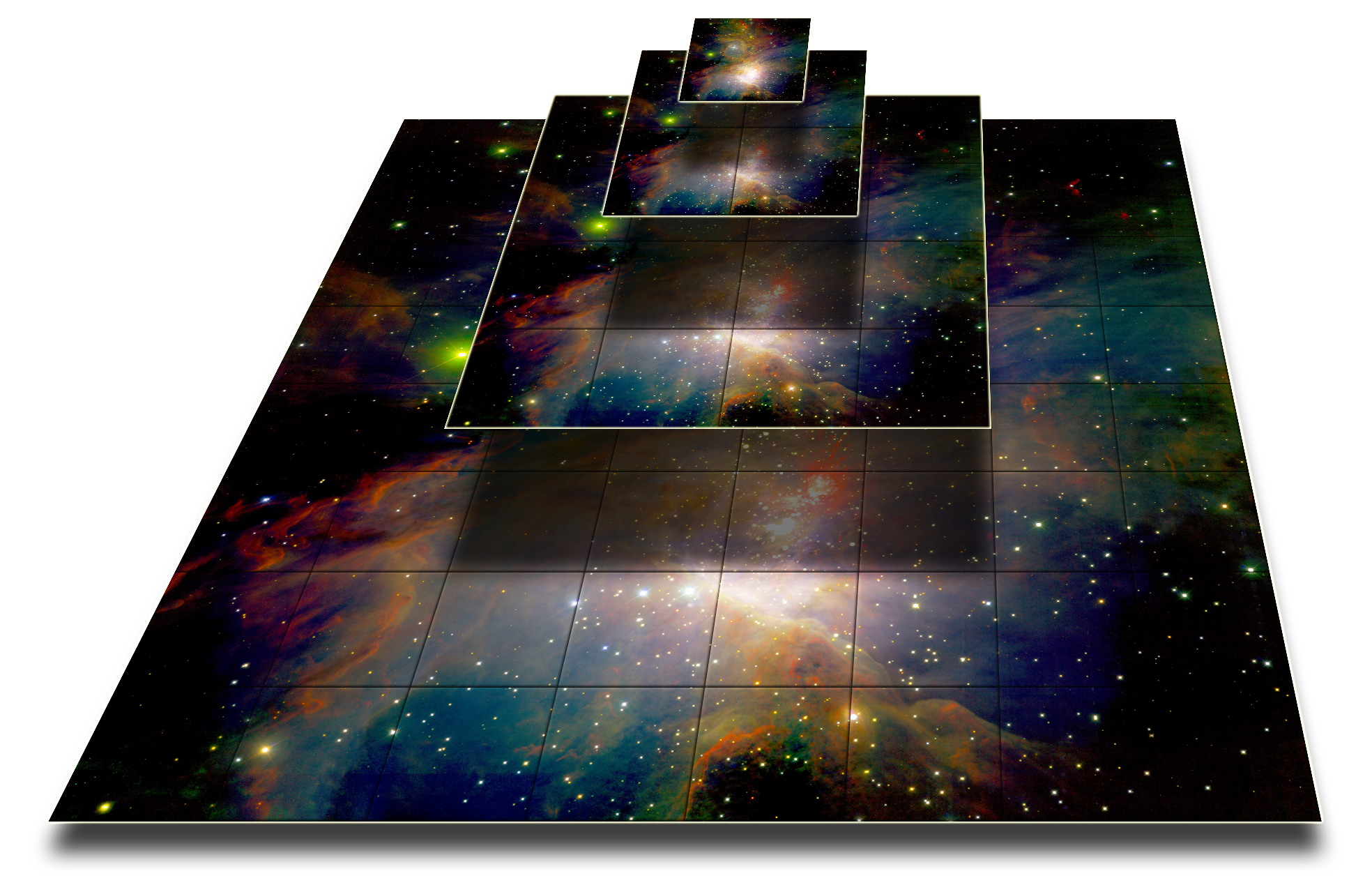}
      \caption{Illustration of tiled multi-resolution pyramid with 4 levels of resolution.}
         \label{fig:pyramid}
   \end{figure}

Transcoding from basic FITS to multi-resolution tiled TIFF is carried out via the {\sc STIFF} conversion package \citep{2012ASPC..461..263B}. The multi-resolution structure consists of an image ``pyramid'' whereby pixels in each image are successively rebinned $2\times2$ and stored in separate TIFF virtual ``directories'' in tiled format. Tile size remains constant across all resolution levels (Fig. \ref{fig:pyramid}). The total number of pixels stored in the pyramid is increased by approximately one third compared to the original raster, but TIFF's widespread support for various lossless compression algorithms (e.g., LZW, Deflate) mitigates some of this extra structural overhead. Note that using pixel rebinning instead of decimation (as in traditional astronomy image display tools) averages out background noise as one zooms out: this makes faint background features such as low surface brightness objects or sky subtraction residuals much easier to spot.

The default orientation for TIFF images (and most image formats) is such that the first pixel resides in the upper left corner of the viewport, whereas FITS images are usually displayed with the first pixel in the lower left corner. To comply with these conventions, {\sc STIFF} flips the original image content along the y direction by proceeding through the FITS file backwards, line-by-line.

{\sc STIFF} takes advantage of the TIFF header ``tag'' mechanism to include metadata that are relevant to the {\sc IIPImage} server and/or web clients. For instance, the {\tt ImageDescription} tag is used to carry a verbatim copy of the original FITS header. Another set of information of particular importance, especially with floating point data, is stored in the {\tt SMinSampleValue} and {\tt SMaxSampleValue} tags: these are the minimum and maximum pixel values ($S_{\rm min}$ and $S_{\rm max}$) that define the display scale. These values do not necessarily represent the full range of pixel values in the image, but rather a range that provides the best visual experience given the type of data. {\sc STIFF} sets $S_{\rm max}$ to the 999${\rm th}$ permil of the image histogram by default. $S_{\rm min}$ is computed in a way that the sky background $S_{\rm sky}$ should appear on screen as a dark grey $\rho_{\rm sky}\approx 0.001$ (expressed as a fraction of the maximum display radiant emittance: $1 \equiv $~full white):
\begin{equation}
S_{\rm min} = \frac{S_{\rm sky} - \rho_{\rm sky}\,S_{\rm max}}
		{1 - \rho_{\rm sky}}.
\end{equation}

{\sc STIFF} currently takes simply the median of all pixel values in the FITS file to compute $S_{\rm sky}$, although better estimates could be computed almost as fast \citep{1996A&AS..117..393B}.

Transcoding speed can be a critical issue, for instance in the context of real-time image monitoring of astronomy observations. On modern hardware, the current {\sc STIFF} conversion rate for transcoding a FITS file to an IIPImage-ready tiled pyramidal TIFF ranges from about 5Mpixel/s to 25Mpixel/s (20-100MB/s) depending on the chosen TIFF compression scheme and system I/O performance. This means that FITS frames with dimensions of up to $~$16k$\times$16k pixels can be converted in a matter of seconds, and just-in-time conversion is a viable option for such images. Note that although {\sc STIFF} is multithreaded, all calls to {\tt libtiff} for writing tiles are done sequentially in the current implementation and there may, therefore, be some room for significant performance improvements.

\subsection{Protocol and Server-Side Features}
{\sc IIPImage} is based on the Internet Imaging Protocol (IIP), a simple HTTP protocol for requesting images or regions within an image, which allows the user to define resolution level, contrast, rotation and other parameters. The protocol was originally defined in the mid-1990s by the \textit{International Imaging Industry Association} \citep{iiprotocol}, but has since been extended for {\sc IIPImage}. The use of such a protocol provides a rich RESTful-like interface to the data, enabling flexible and consistent access to imaging data. {\sc IIPImage} is also capable of communicating using the simpler tile request protocols used by Zoomify or Deepzoom and the more recent IIIF image access API \citep[International Image Interoperability Framework,][]{iiif}.

Table \ref{tab:commands} lists the main commands already available in the original, cultural heritage-oriented version of {\sc IIPimage}. For a complete description of the protocol, see the full IIP protocol specification \citep{iiprotocol}.

\setlength{\tabcolsep}{8pt}
\renewcommand{\arraystretch}{1.5}
\begin{table}[h]
\begin{small}
\begin{tabular}{|p{0.15\columnwidth}|p{0.72\columnwidth}|}
\hline
\textbf{Command} & \textbf{Description}\\
\hline
\hline
{\tt FIF} & Image path $p$. [\texttt{FIF=p}]\\
\hline
{\tt OBJ} & Property/ies $text$ to be retrieved from image and server metadata. [\texttt{OBJ=text}]\\
\hline
{\tt QLT} & JPEG quality factor $q$ between 0 (worst) and 100 (best). [\texttt{QLT=q}]\\
\hline
{\tt SDS} & Specify a particular image within a set of sequences or set of multi-band images. [\texttt{SDS=s1,s2}]\\
\hline
{\tt CNT} & Contrast factor $c$. [\texttt{CNT=c}]\\
\hline
{\tt CVT} & Return the full image or a region, in JPEG format. [\texttt{CVT=jpeg}]\\
\hline
{\tt WID} & Width $w$ in pixels of the full sized JPEG image returned by the {\tt CVT} command (interpolated from the nearest resolution). [\texttt{WID=w}]\\
\hline
{\tt HEI} & Height $h$ in pixels of the full sized JPEG image returned by the {\tt CVT} command (interpolated from the nearest resolution). [\texttt{HEI=h}]\\
\hline
{\tt RGN} & Define a region of interest starting at relative coordinates $x$, $y$ with width $w$ and height $h$. [\texttt{RGN=x,y,w,h}]\\
\hline
{\tt ROT} & Rotate the image by $r$ (90, 180 or 270 degrees). [\texttt{ROT=r}]\\
\hline
{\tt JTL} & Return a tile with index $n$ at resolution level $r$, in JPEG format. [\texttt{JTL=r,n}] \\
\hline
{\tt SHD} & Apply hillshading simulation with azimuth and altitude angles $a$, $b$. [\texttt{SHD=a,b}]\\
\hline
{\tt SPECTRA} & Return pixel values in all image channels for a particular point $x$,$y$ on tile $t$ at resolution $r$ in XML format. [\texttt{SHD=r,t,x,y}]\\
\hline
\end{tabular}
\end{small}
\caption{
\label{tab:commands}
Main commands available in {\sc IIPImage}
}
\end{table}

For this project the entire {\sc IIPImage} codebase was updated and generalized to handle up to 32 bits per pixel, with support for single precision floating point data. Support for double precision (at the expense of performance) would require a relatively simple code update. In addition, several extensions were implemented that allow the application of predefined colormaps to grayscale images, adjust the gamma correction, change the minimum and maximum cut-offs of the pixel value range, and that enable the export of image data profiles. A list of the new available commands is given in table \ref{tab:newcomm}.

\begin{table}[h]
\begin{small}
\begin{tabular}{|p{0.15\columnwidth}|p{0.72\columnwidth}|}
\hline
\textbf{Command} & \textbf{Description}\\
\hline
\hline
{\tt CMP} & Set the colormap for grayscale images. Valid colormaps include
{\tt GREY}, {\tt JET}, {\tt COLD}, {\tt HOT}, {\tt RED}, {\tt GREEN} and {\tt BLUE}. [\texttt{CMP=JET}]\\
\hline
{\tt INV} & Invert image or colormap. [\texttt{Does not require an argument}] \\
\hline 
{\tt GAM} & Set gamma correction to $g$. [\texttt{GAM=g}] \\ 
\hline 
{\tt MINMAX} & Set minimum $min$ and maximum $max$ for channel $c$. [\texttt{MINMAX=c:min,max}]\\ 
\hline
{\tt PFL} & Request full bit-depth data profile for resolution $r$ along the line joining pixel $x1$,$y1$ to $x2$,$y1$. [\texttt{PFL=r:x1,y1-x2,y2}]

{\it\footnotesize Note: Only horizontal ($y1=y2$) and vertical profiles ($x1=x2$) currently supported}\\
\hline

\end{tabular}
\end{small}
\caption{
\label{tab:newcomm}
List of new commands implemented in {\sc IIPImage}.
}
\end{table}

\subsubsection{Examples}
\label{subsec:examples}

In order to better understand how these commands can be used, here are several examples showing the typical syntax and usage for applying colormaps, setting a gamma correction and for obtaining a full bit-depth profile.

All requests take the general form:
\begin{lstlisting}
<protocol>://<server address>/<iipsrv>?<IIP Commands>
\end{lstlisting}

The first IIP command must specify the image path and several IIP command--value pairs can be chained together using the separator $\&$ in the following way:
\begin{lstlisting}
FIF=<image path>&<command>=<value>&<command>=<value>
\end{lstlisting}

Thus, a typical request for the tile that fits into the smallest available resolution (tile 0 at resolution 0) of a TIFF image named \texttt{image.tif} is:

\begin{lstlisting}
http://server/iipsrv.fcgi?FIF=image.tif&JTL=0,0
\end{lstlisting}

Let us now look at some more detailed examples using the new functionality created for {\sc IIPImage}. For example, in order to export a profile in JSON format from pixel location $x_{\rm 1}$,$y_{\rm 1}$ horizontally to pixel location $x_{\rm 2}$,$y_{\rm 2}$ at resolution $r$ on image \texttt{image.tif}, the request would take the form:

\begin{lstlisting}
FIF=image.tif&PFL=r:x1,y1-x2,y2
\end{lstlisting}

In order to request tile $t$ at resolution $r$ and apply a standard \textit{jet} colormap to image \texttt{image.tif}, the request would take the form:

\begin{lstlisting}
FIF=image.tif&CMP=JET&JTL=r,t
\end{lstlisting}

and the equivalent inverted colormap request:
\begin{lstlisting}
FIF=image.tif&CMP=JET&INV&JTL=r,t
\end{lstlisting}

In order to obtain metadata containing the minimum and maximum values per channel:
\begin{lstlisting}
FIF=image.tif&OBJ=min-max-sample-values
\end{lstlisting}

In order to request tile $t$ at resolution $r$ and apply a gamma correction of $g$ and specify a minimum and maximum of $m_{\rm 1}$ and $m_{\rm 2}$ respectively for image band $b$:

\begin{lstlisting}
FIF=image.tif&MINMAX=b:m1,m2&GAM=g&JTL=r,t
\end{lstlisting}

Commands are not order sensitive excepting {\tt JTL} and {\tt CVT} that must always be specified last.

\subsection{Security}
A client-server architecture also has the advantage in terms of control and security of the data as the raw data at full bit depth does not necessarily need to be made fully available to the end user. Indeed, the raw data need never be directly accessible by the public and can be stored on firewalled internal storage and only accessible via the {\sc IIPImage} server. Thus only 8 bit processed data is ever sent out to the client and restrictions and limits can be applied if fully open access is not desired. The {\sc IIPImage} server also contains several features for added security, such as a path prefix, which limits access to a particular subdirectory on the storage server. Any requests to images higher up or outside of this subdirectory tree are blocked.

If an even greater level of security is required on the transmitted data, the {\sc IIPImage} server can also dynamically apply a watermark to each image tile with a configurable level of opacity. Watermarking can be randomized both in terms of which tiles they are applied to as well as their position within the tile itself, making removal of watermarks extremely difficult.

\subsection{Web Clients}
\label{chap:client}
Two web clients, developed using different approaches and different goals in mind, are presented in this paper as examples to illustrate the capabilities of the system.

The first one, known as {\sc VisiOmatic}, is built on top of the {\sc leaflet} JavaScript mini-framework, and is designed to display large celestial images through a classic image tile-based view.

The second client builds on the existing {\sc IIPMooViewer} client to demonstrate two experimental features more specifically relevant to planetary surface studies: hillshading and advanced compositing / filtering performed at the pixel level within the browser.

\section{Astronomy Applications}
\label{chap:astrapp}

\subsection{Celestial Images}

Two essential features of astronomy image browsers are missing in the IIPMooViewer client originally developed for cultural heritage applications: the handling of celestial coordinates and a comprehensive management system for vector layers (overlays). It soon became clear that developing such a system from scratch with limited human resources would raise severe maintenance issues and portability concerns across browsers and platforms. We investigated several JavaScript libraries that would provide such functionality and decided to build a new client, {\sc VisiOmatic}, based on the {\sc Leaflet} library \citep{leafletjs}. {\sc Leaflet} is open-source and provides all the necessary functions to build a web client for browsing interactive maps. It is, in fact, not simply a client, but a small framework, offering features not directly available in standard JavaScript such as class creation and inheritance. It has a well-documented, user-friendly API and a rich collection of plug-ins that significantly boost its potential, while providing many advanced programming examples. Indeed, {\sc VisiOmatic} operates as a {\sc Leaflet} plug-in and as such comes bundled as a NodeJS package. Documentation for the {\sc VisiOmatic} API is available on the {\sc VisiOmatic} {\sc GitHub} page\footnote{\url{https://github.com/astromatic/visiomatic}}.

Once the {\tt iipsrv} server has been installed, embedding a zoomable astronomy image in a web page with the {\sc VisiOmatic} and {\sc Leaflet} JavasScript libraries is very simple and can be done with the following code:

\begin{lstlisting}
 <div id="map"></div>
  <script>
   var map = L.map('map'),
       layer = L.tileLayer.iip('/fcgi-bin/iipsrv.fcgi?FIF=/path/to/image.ptif').addTo(map);
  </script>
\end{lstlisting}

{\sc Leaflet} was built from the ground up with mobile device support in mind. {\sc VisiOmatic} capitalizes on this approach by defining the current map coordinates at the center of the viewport instead of the mouse position. This also makes the coordinate widget display area usable for input, copy or paste as coordinates do not change while moving the mouse. Celestial coordinates are handled through a custom JavaScript library that emulates a small subset of the WCS standard \citep{2002A&A...395.1077C}, based on the FITS header content transmitted by the {\sc IIPImage} server. Our simplified WCS library fits into {\sc Leaflet}'s native latitude--longitude coordinate management system, giving access to all layer contents directly in celestial coordinates. This makes it particularly easy to synchronize maps that do not use the same projection for e.g., orientation maps, ``smart'' magnifiers, or multi-band monitoring.

Changing image settings is done by appending the relevant IIP commands (see e.g., Table \ref{tab:newcomm}) to the {\tt http} {\tt GET} tile requests. Metadata and specific data queries, such as profile extractions, are carried out through AJAX requests. {\sc VisiOmatic} also uses AJAX requests for querying catalogs from other domains, with the restriction that the {\it same origin security policy}\footnote{The {\it Cross-Origin Resource Sharing} (CORS) mechanism implemented in modern browsers could in principle prevent that, but it is not supported by the main astronomy data providers at this time.} present in current browsers requires that all requests transit through the image server domain, which must, therefore, be configured as a web proxy.

\begin{figure*}[htb!]
   \centering
   \includegraphics[width=\textwidth]{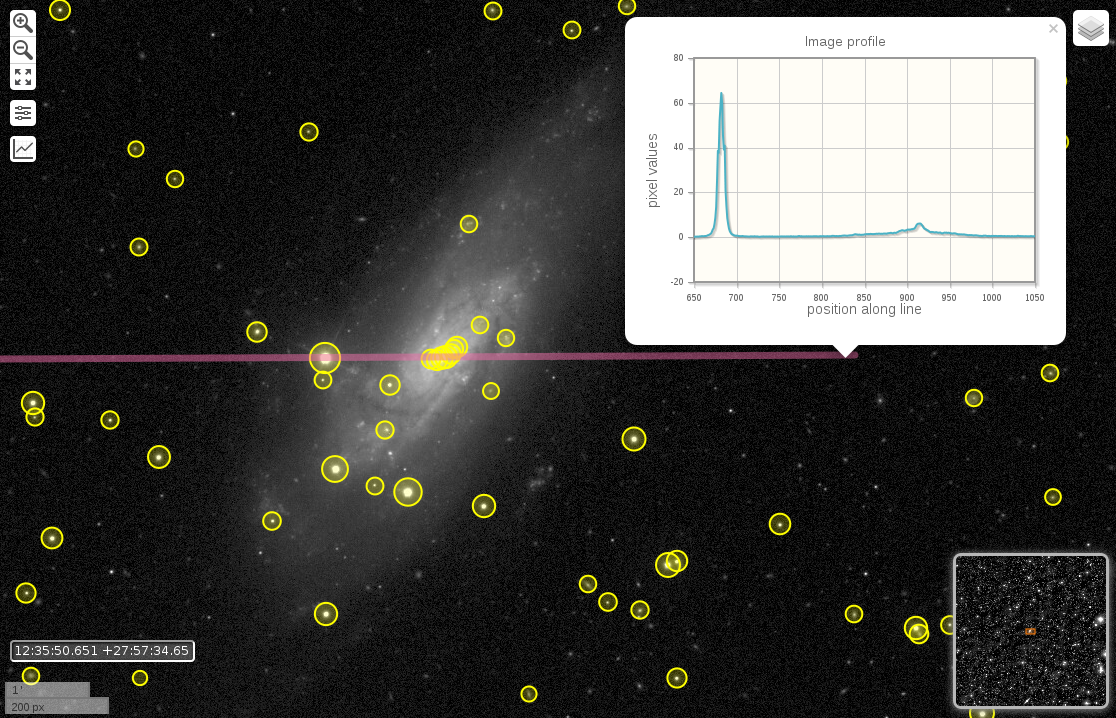}
      \caption{The Visi{\it O}matic web client showing a part of an SDSS release nine image stack \citep{2012ApJS..203...21A} provided by the {\sc IIPImage} server in the main layer, plus two vector layers superimposed. Yellow: local detections from the photometric SDSS catalog provided by the Vizier service \citep{2000A&AS..143...23O}. Purple: horizontal profile through the image extracted by the {\sc IIPImage} server.}
         \label{fig:visiomatic}
   \end{figure*}

The {\sc VisiOmatic} website\footnote{\url{http://visiomatic.org}} showcases several examples of applications built with the {\sc VisiOmatic} client. They involve large images of the deep sky stored in floating point format, including a one terabyte (500,000 $\times $500,000 pixels) combination of 250,000 exposures from the 9$^{\rm th}$ Sloan Digital Sky Survey data release \citep{2012ApJS..203...21A}, representing about 3TB worth of raw image data (Fig. \ref{fig:visiomatic}).

Display performance with the {\sc VisiOmatic} client varies from browser to browser. Browsers based on the {\sc WebKit} rendering engine (e.g., {\sc Chrome}, {\sc Safari}) generally offer the smoothest experience on all platforms, especially with complex overlays.
User experience may also vary because of the different ways browsers are able to deal with data. For example, examining images at exceedingly high zoom levels and scrutinizing groups of pixels displayed as blocks is common practice among astronomers. {\sc Leaflet} takes advantage of the built-in resampling engines in browsers to allow image tiles to be zoomed in smoothly through CSS3 animations. {\sc VisiOmatic} uses the {\tt image-rendering} CSS property to activate nearest-neighbor interpolation and have the pixels displayed as blocks at higher zoom levels. Although this works in, for example, {\sc Firefox} and {\sc Internet Explorer 11}, other browsers, such as {\sc Chrome}, do not offer the possibility to turn off bilinear interpolation at the present time, and zoomed images will not appear pixelated in those browsers. Hopefully, it is expected that such residual differences will eventually disappear as browser technology converges over standards.

\section{Planetary Science}
\label{chap:planetapp}
Planetary Science data are largely heterogeneous with respect to the physical quantities they describe (chemical abundances, atmospheric composition, magnetic and gravitational fields of Earth-like planets and satellites, reflectance, surface composition) and with respect to the formats they are encoded in (raster, vector, time-series, in ASCII or various binary formats). Two scientific communities are essentially involved in Planetary Science research: astronomers and geologists / geophysicists.

Geographical Information Systems (GIS) are the basis for planetary surface studies but they often suffer from a lack of systematic and controlled access to pixel values for quantitative physical analyses on raster data \citep{2012ASPC..461..411M}.

In Earth Sciences, distributors of GIS commercial software have been ready to exploit the potential of the Web. This is the case, for example, of the online ArcGIS WebMap Viewer\footnote{\url{http://www.arcgis.com/home/webmap/viewer.html}}.

However, the difficulty in sending 16 or 32 bit precision scientific data using current web technologies has, hitherto, limited web visualization to public outreach applications such as the Microsoft World Wide Telescope available for images of Mars \citep{2011LPI....42.2337S} or Google Earth for Mars\footnote{\url{http://www.google.com/earth/explore/showcase/mars.html}}.

Nevertheless, remote scientific visualization has been achieved with tools such as JMars\footnote{\url{http://jmars.mars.asu.edu}} \citep{2009AGUFMIN22A..06C} and HiView\footnote{\url{http://hirise.lpl.arizona.edu/hiview/}}, which are both Java clients that aim to visualize both remote and local data. The first is GIS based (layer superposition oriented) while HiView is more a remote sensing software (raster manipulation oriented) that is an ad-hoc product for HiRISE\footnote{\url{http://hirise.lpl.arizona.edu}} and which uses the JPIP protocol for remote access to JPEG2000 imagery.

The visualization system we propose already supports basic manipulations on raster layers, raster layer superposition and could easily manage vector layer creation and superposition. It, moreover, enables access to full precision pixel values and provides a simple and generic solution for planetary applications, efficiently and elegantly blending both GIS and remote sensing approaches.

\subsection{Color Compositing}
\begin{figure}
   \centering
   \includegraphics[width=\columnwidth]{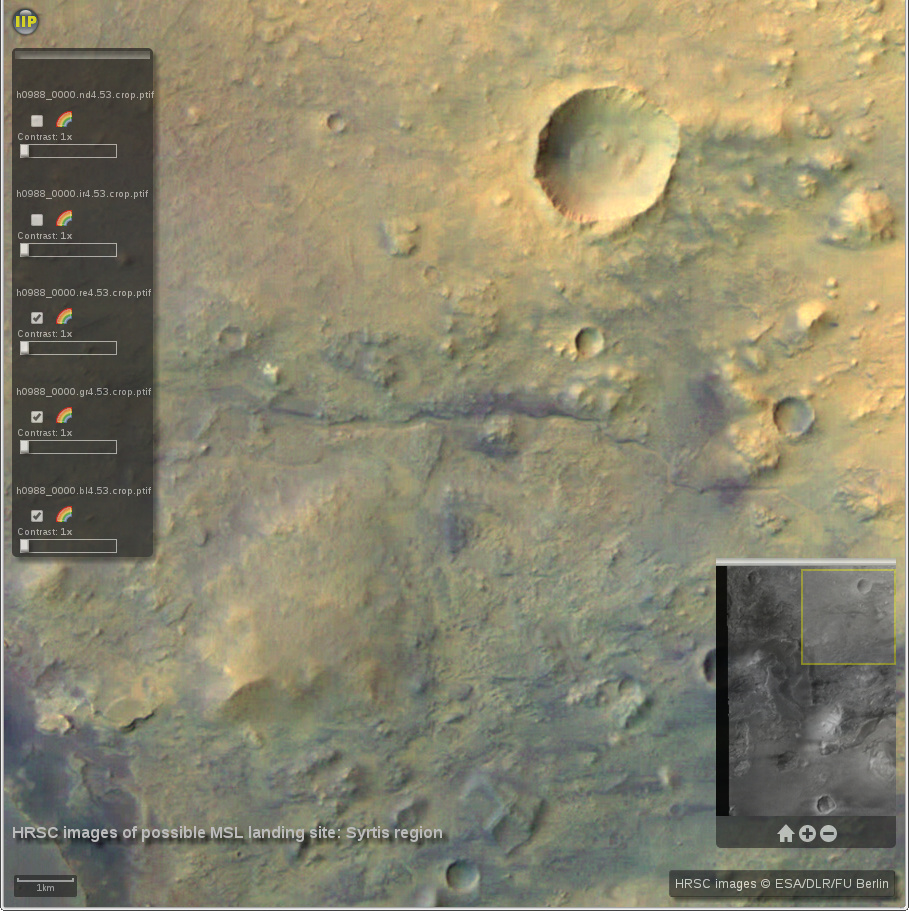}
      \caption{Example of a planetary application using HRSC Mars images (ESA/DLR/FU Berlin/G. Neukum). The resulting color image is a linear combination of input channels. The mixing matrix is defined by a user-adjustable combination of Red, Green, Blue and a contrast factor for each input channel.}
         \label{fig:hrsc}
   \end{figure}

Color compositing is an essential feature in both GIS and remote sensing applications and is used to point out differences in surface composition by performing on-demand composition of specific color bands. Interactive color composition on the Web can be achieved using the HTML5 \texttt{canvas} element which allows us to directly access and manipulate image pixel values. This, therefore, enables more complex real-time image processing directly within the client and we have developed a version of our client making extensive use of HTML5 \texttt{canvas} properties\footnote{\url{http://image.iap.fr/iipcanvas/hrsc.html}} in order to implement on-demand color composition with multiple input channels (Fig. \ref{fig:hrsc}).

Color compositing performance depends essentially on canvas size (the overall image size is irrelevant as only the displayed part of the image needs to be processed). For the example cited above the processing time is about 1 to 3 ms per tile ($256\times 256$ pixels), depending on browser and client hardware.

\subsection{Terrain Maps - Hillshading}

High resolution 3D data is not easy to stream or to make multi-resolution. Furthermore as {\sc IIPImage} is essentially image-centric, a 2D rendering approach to visualization was favored. In order to facilitate the use of DEM (digital elevation map) data, two approaches have been developed in our {\sc IIPimage} framework.

The first approach is the dynamic application of custom colormaps to grayscale images. A new command \texttt{CMP} has been added to the IIP protocol, which can also be useful for the visualization of other physical map data such as gas density, temperature or chemical abundances in real or simulated data. 

\begin{figure*}[htb!]
   \centering
   \includegraphics[width=\textwidth]{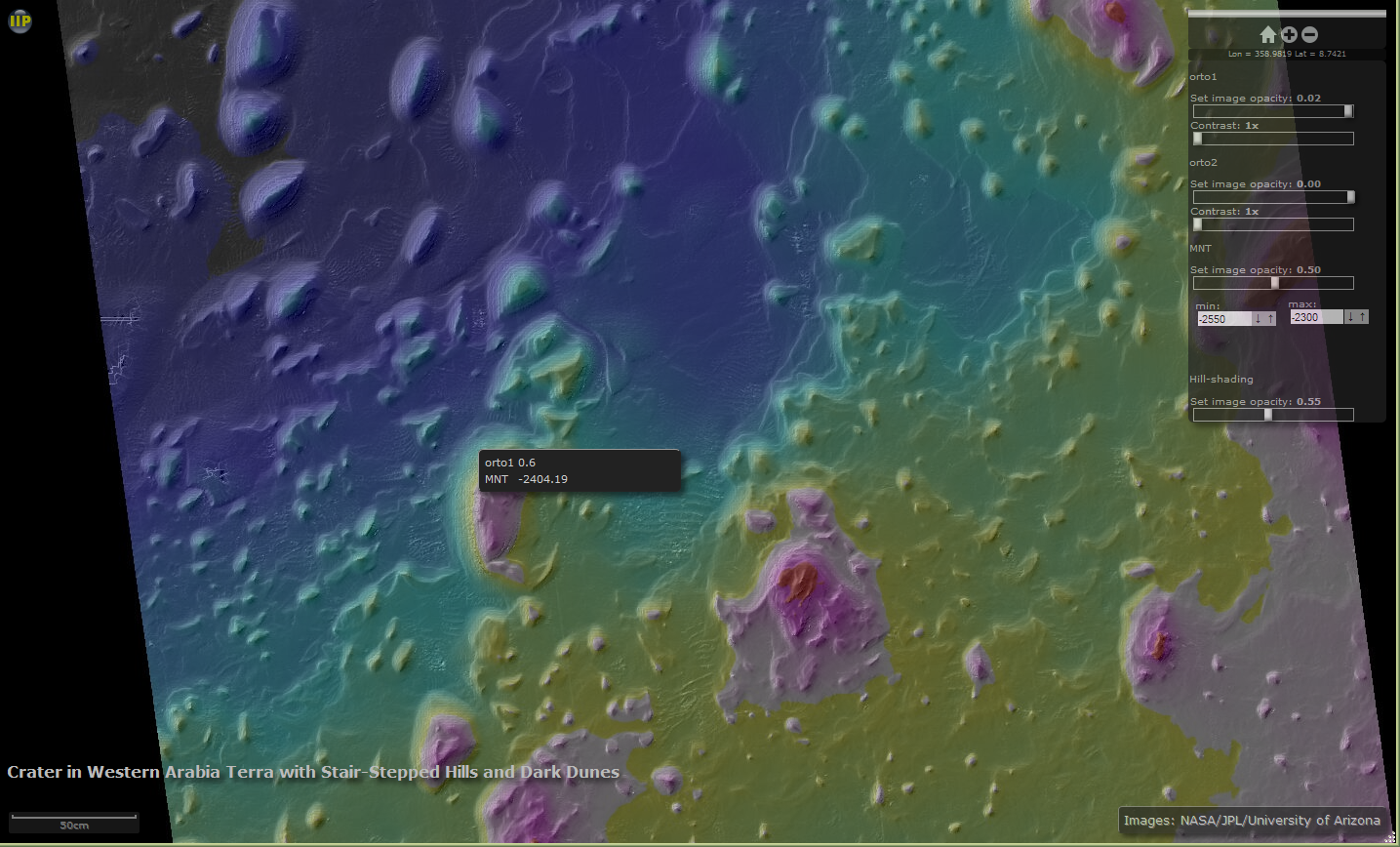}
      \caption{Example of a planetary web application using HiRISE Mars images (NASA/JPL/University of Arizona). The digital elevation model (a floating point raster) is displayed using the JET colormap with cuts set by the user from the control panel. Superimposed is the hill-shading layer computed by the {\sc IIPImage} server from the DEM; the azimuth incidence angle can be adjusted from the control panel.}
         \label{fig:hirise}
   \end{figure*}

In the second approach, elevation point data is converted to vector normal and height data at each pixel. In this form, they are also able to be stored within the standard TIFF format. The $XYZ$ normal vectors can be packed into a 3 channel ``color'' TIFF, whereas the height data can be packed into a separate 1 channel monochrome TIFF. They are both, therefore, able to be tiled, compressed and structured into a multi-resolution pyramid for streaming with {\sc IIPImage}.
A basic rendering technique for DEM data is that of hill-shading \citep{horn_hill_1981} where a virtual directional illumination is used to create shading on virtual ``hills''. A fast hill-shading algorithm has been implemented server-side in {\sc IIPImage} and extended to 32 bit data allowing the user to interactively set the angle of incidence of the light source and view a dynamically rendered hill-shaded relief map. An example showing a Mars terrain map from Western Arabia Terra can be seen online\footnote{\url{http://image.iap.fr/iipdiv/hirise.html}} and in Fig. \ref{fig:hirise}.

\section{Performance Analysis}
\label{chap:perf}
Although the use of JPEG compression as a delivery format significantly reduces the bandwidth required, data access, dynamic processing and compression of 32 bit data can impose significant server-side overhead, that will ultimately dictate the maximum number of users that a server will be able to handle.
Timings and memory usage depend on image type, server settings and commands in the query; we chose to focus on the typical case of browsing a large, single channel, single precision floating-point image stored with tiles of $256\times256$ pixels in size. In order to fully test this, we created a large $131,072\times 131,072$ pixel FITS image by combining contiguous SDSS i-band images using the {\sc SWarp} package \citep{2002ASPC..281..228B}. This large image was then converted to a 92GB multi-resolution TIFF comprising 9 resolution levels using {\sc STIFF}.

Our tests were performed on two Dell PowerEdge servers running GNU/Linux (Fedora distribution with kernel 3.11) and equipped with 2.6GHz processors, 32 and 48 GB of RAM and a Perc5i internal RAID controller. In order to check the influence of the I/O subsystem on server performance, we installed the TIFF file on two different types of RAID:

\begin{itemize}
\item[a)] a RAID 6 array of $12\times3$ TBytes SAS (6Gb/s) hard drives formatted with the XFS filesystem.
\item[b)] a RAID 5 array of $6\times1$ Tbytes SATA3 (6Gb/s) solid-state drives (SSDs) formatted with the Ext4 filesystem.
\end{itemize}

On both systems we obtain a typical sequential read speed of 1.2 GB/s for large blocks; but obviously access times are much lower on the RAID of SSDs ($<1$ms vs 15ms for the one with regular hard drives).

The client consists of a third machine sending requests to any of the two servers through a dedicated 10GbE network. We used a modified version of {\sc ab}, the {\sc ApacheBench} HTTP server benchmarking package, to send sequences of requests to random tiles among the 262,144 that compose the highest image zoom level. Appropriate system settings, as prescribed in \cite{Veal:2007:PSM:1323548.1323562}, were applied server-side and client-side to ensure that both ends would stand the highest possible concurrency levels with minimum latencies and maximum throughput. We conducted preliminary tests through Apache's httpd\footnote{\url{http://httpd.apache.org}}, Lighty Labs' lighttpd\footnote{\url{http://www.lighttpd.net}}, a combination of Nginx\footnote{\url{http://nginx.org}} and Lighty Labs' {\tt spawn-fcgi} and finally LiteSpeed Technologies' OpenLiteSpeed\footnote{\url{http://open.litespeedtech.com}}. We found the latter to offer the best combination of performance and robustness, especially at high concurrency levels; hence all the requests to {\tt iipsrv} in the tests reported below were served through OpenLiteSpeed (one single {\tt lshttpd} instance).

\subsection{Timings}
\begin{figure*}[htb]
   \centering
   \includegraphics[width=0.49\textwidth]{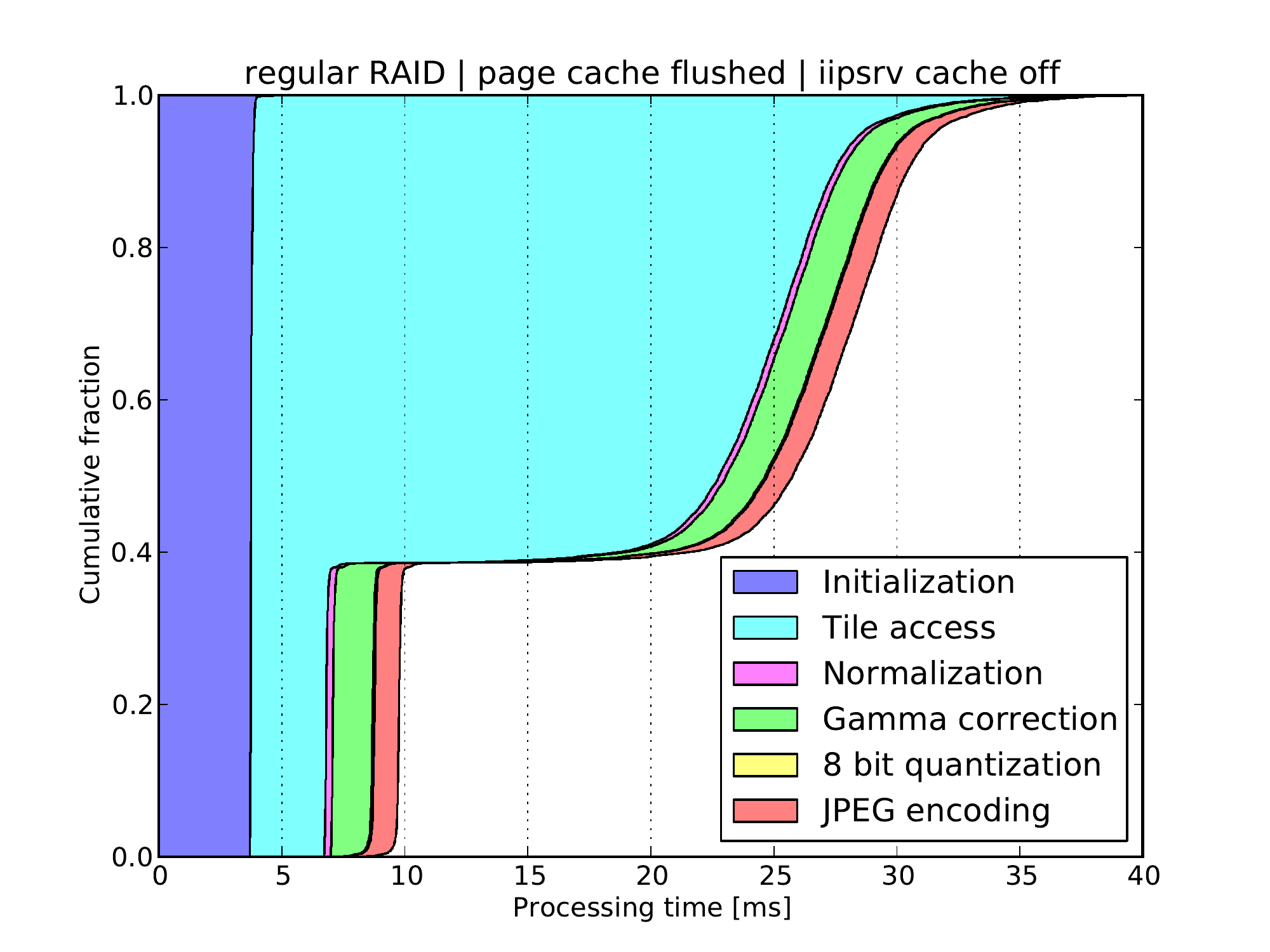}
   \includegraphics[width=0.49\textwidth]{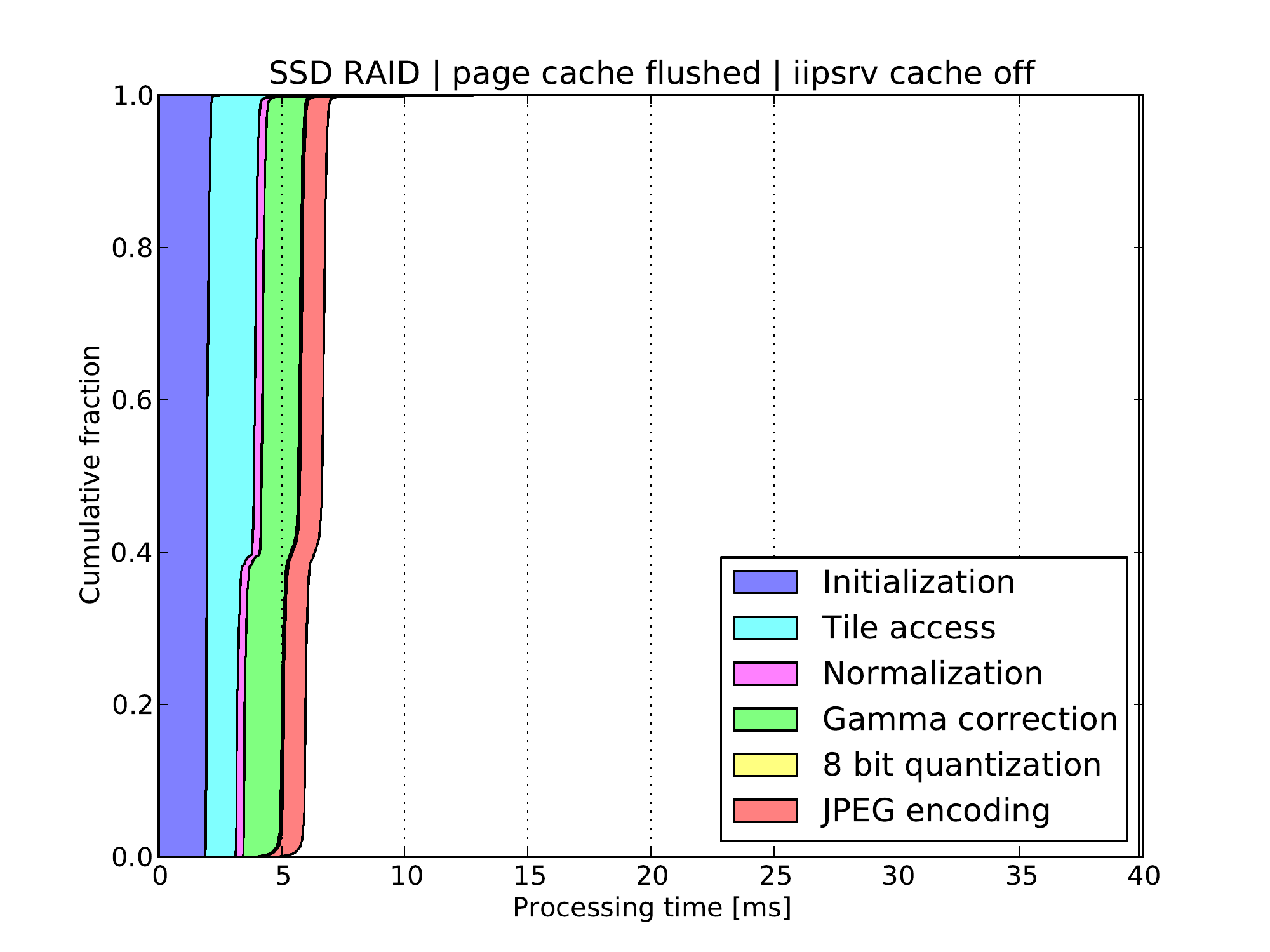}
   \centering
   \includegraphics[width=0.49\textwidth]{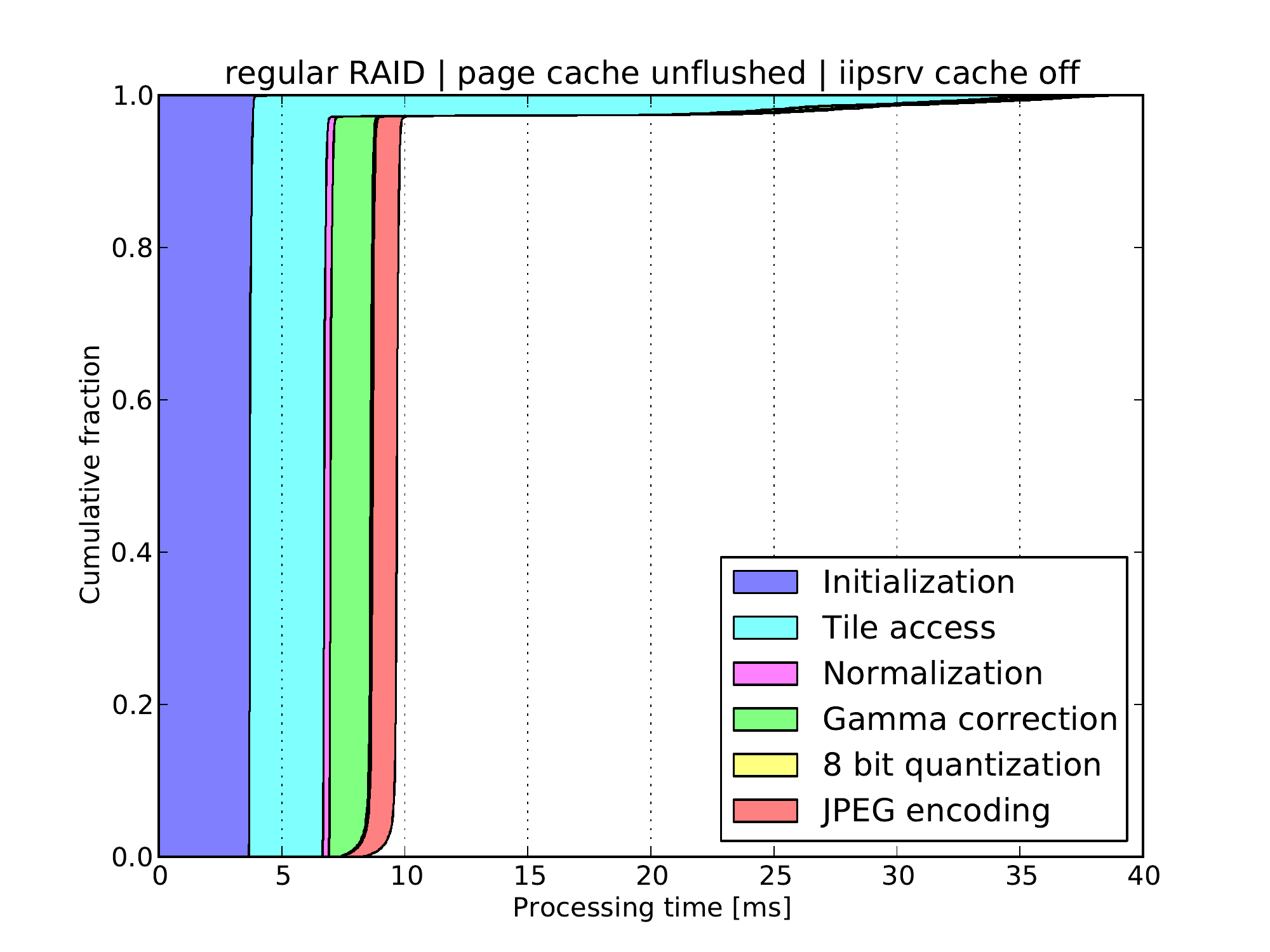}
   \includegraphics[width=0.49\textwidth]{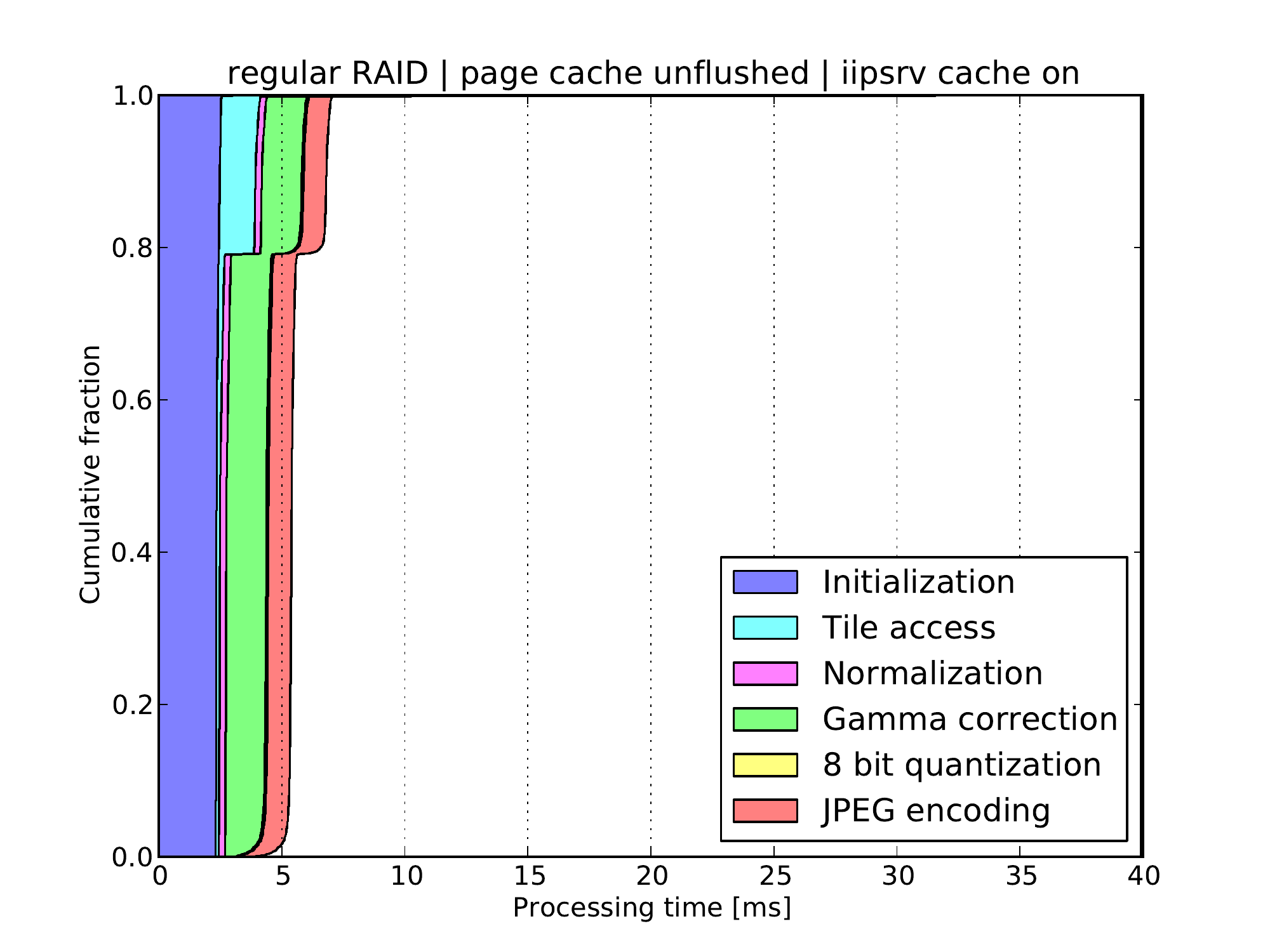}
      \caption{Cumulative distributions for the timings of the main tasks involved in the processing of random $256\times256$ pixel tiles in four different contexts (see text for details). ``Initialization'' is the time taken to initialize various objects and (re-)open the TIFF file that contains the requested raw tile (call to {\sc TIFFOpen()}). ``Tile access'' is the time spent accessing and reading the content of a raw tile. ``Normalization'' and ``Gamma correction'' respectively measure the time it takes to apply intensity cuts and to compress the dynamic range of pixel values for the whole tile. ``8 bit quantization'' is the time spent converting the tile to 8 bit format, while ``JPEG encoding'' is the time taken to encode the tile in JPEG format.}
     \label{fig:timings}
   \end{figure*}

Figure \ref{fig:timings} shows the distribution of timings of the main tasks involved in the server-side processing of a tile, for several system and {\tt iipsrv} cache settings. In order to probe the impact of I/O latencies, we set up an experiment where the server system page cache is flushed and the {\tt iipsrv} internal cache is deactivated prior to running the test (upper row of Fig. \ref{fig:timings}). With such settings most accesses to TIFF raw tiles do not benefit from caching. As a consequence, {\tt iipsrv} timings are dominated by random file access times when the data are stored on spinning hard disks, with access latencies reaching up to $\approx 25$ms in unfavorable situations. As expected, switching to SSDs reduces the uncached file access latencies to less than 1ms.

However, in practice much better timings will be obtained with regular hard drives, as tiles are generally not accessed randomly. Moreover, leaving the system page cache un-flushed between test sessions when using spinning hard drives reduces access latencies to a few milliseconds (lower row in Fig. \ref{fig:timings}). Activating {\tt iipsrv}'s internal cache system will further reduce latencies close to zero for tiles that were recently visited.

Further testing with TIFF images of different size was carried out in order to ensure that the system would also scale in terms of file size and the timings reported above remain roughly identical as file size increases up to at least 1.8TB.

\subsection{Concurrency and Data Throughput}
Each single-threaded FastCGI process takes about 5-10ms to complete, and is therefore capable of serving up to 100-200 $256\times 256$ ``new'' tiles per second. Higher tile serving rates are obtained by running several instances of {\tt iipsrv} on servers with multiple CPU cores. But how is the system able to keep up with a large number of concurrent requests?

\begin{figure}[htb!]
   \centering
   \includegraphics[width=\columnwidth]{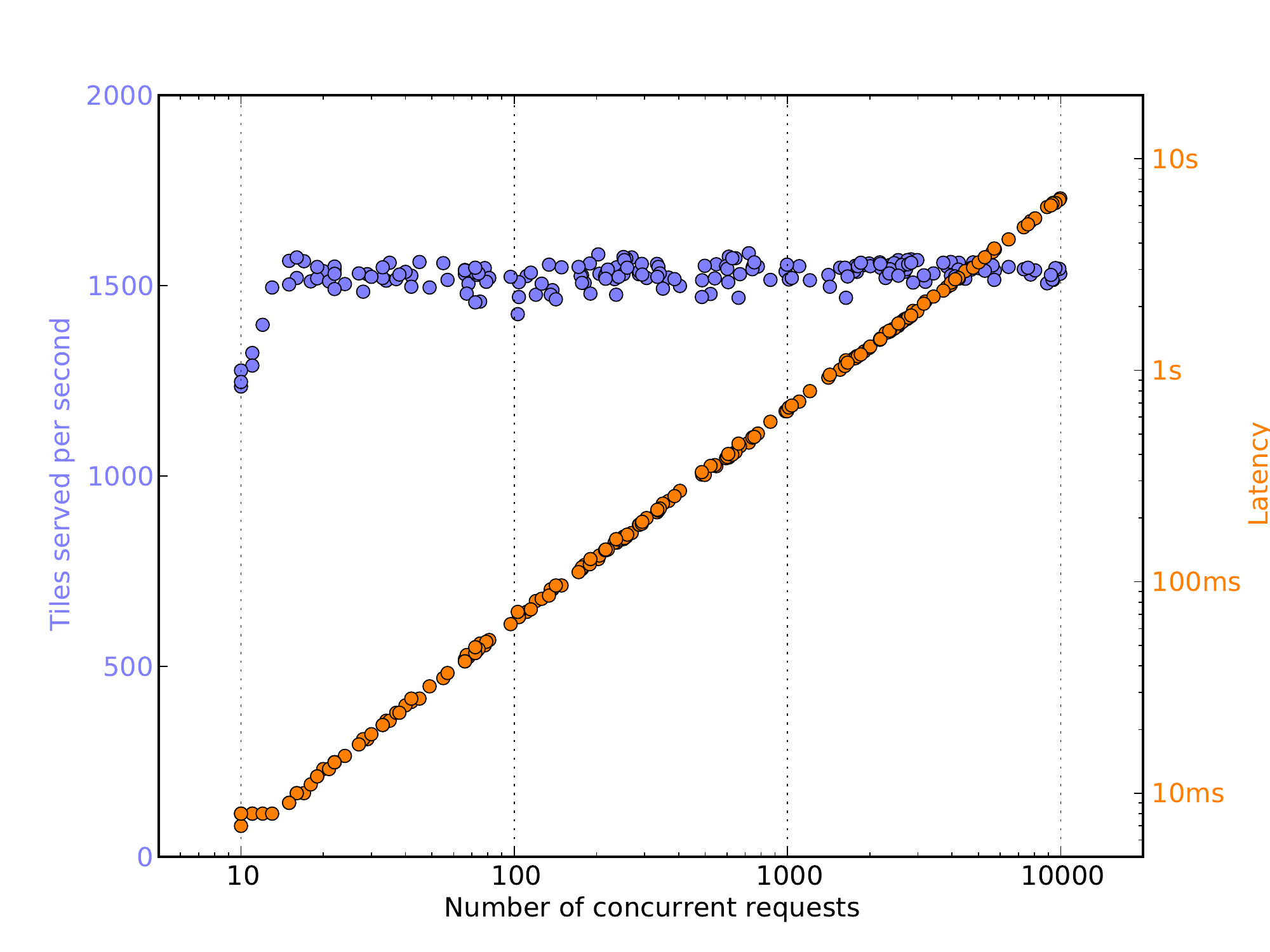}
      \caption{Tile serving rate (in blue) and latency (in orange) as a function of the number of concurrent tile requests using 12 instances of {\tt iipsrv} on a 12-core server equipped with an SSD RAID.}
     \label{fig:concur}
   \end{figure}

As Fig. \ref{fig:concur} shows, the tile serving rate remains remarkably flat, and latency scales linearly with the concurrency level when the number of concurrent requests exceeds that of CPU cores. Setting a limit for average latency to $\sim$500ms for comfortable image browsing, we see that a single 12-core web server can handle $\sim$700 concurrent $256\times256$ tile requests, which corresponds to about 100 users frantically browsing large, uncompressed, single-channel floating-point images. This estimation is well verified in practice, although it obviously depends on tile size and on the amount of processing carried out by {\tt iipsrv}. Note that the average tile serving rate obtained with a single 12-core web server corresponds to a sustained data rate of 60MB/s for $256\times256$ tiles encoded at a JPEG quality factor of 90; higher JPEG quality factors bring the data rate close to the saturation limit of a 1GbE connection.

\section{Conclusion and Future Work}
\label{chap:conclu}

A high performance web-based system for remote visualization of full resolution scientific grade astronomy images and data has been developed. The system is entirely open-source and capable of efficiently handling full resolution 32 bit floating point image and elevation map data.

We have studied the performance and scalability of the system and have shown that it is capable of handling terabyte-size scientific-grade images that can be browsed comfortably by at least a hundred simultaneous users, on a single server.

By using and extending an existing open source project, a system for astronomy has been put together that is fully mature, that will benefit from the synergies of the wider scientific imaging community and that is ready for use in a busy production environment. In addition the {\sc IIPImage} server, is distributed as part of the default Debian, Ubuntu and Fedora package repositories, making installation and configuration of the system very straightforward. All the code developed within this project for {\tt iipsrv} has been integrated into the main code base and will form an integral part of the 1.0 release.
However, there are still many potential directions for improvements, both server-side and client-side. Most importantly:

\begin{itemize}
\item {The TIFF storage format used on the server currently restricts pixel bit depth, the number of image channels, and I/O performance (through {\tt libtiff}). A valid alternative to TIFF could be the Hierarchical Data Format Version 5 (HDF5) \citep{the_hdf_group_hierarchical_2000}, which provides a generic, abstract data model that enables POSIX-style access to data objects organized hierarchically within a single file; some radio-astronomers have been trying to promote the use of HDF5 for storing massive and complex astronomy datasets \citep{2012ASPC..461..871M}. A more radical approach would be to adopt JPEG2000 as the archival storage format for astronomy imaging archives \citep{2014arXiv1403.2801K}, which could also remove the need for transcoding images for visualization purposes.}

\item {Additional image operations could be implemented within {\tt iipsrv}, including real-time hyperspectral image processing and compositing.}

\item {Although the {\sc IIPImage} image tile server already supports simple standard tile query protocols and interfaces easily with most image panning clients, a welcome addition would be to offer support for the more GIS-oriented WTMS (Web Map Tile Service) protocol \citep{wmts}}.

\item {The International Virtual Observatory Alliance (IVOA) has agreed on a standard set of specifications for discovering and accessing remote astronomical image datasets: the Simple Image Access Protocol (SIAP) \citep{2011arXiv1110.0499T}. The response to an SIAP query consists of metadata and download URLs for matching image products. Current SIAP specifications\footnote{http://www.ivoa.net/documents/SIA/} do not provide specific ways to access pyramids of tiled images. Still, support for SIAP could be implemented within or outside of {\tt iipsrv} for generating, for example, JPEG cutouts or lists of tiles that match a given set of coordinates/field of view/pixel scale.}

\item {Both {\sc IIPMooViewer} and {\sc Leaflet} clients require all layers displayed on a map at the same moment to share the same ``native'' pixel grid (projection). Although this limitation does not prevent ``blinking'' images with different pixel grids, it precludes overlapping different observations/exposures on screen. For instance it makes it impossible to display accurately the entire focal plane of a mosaic camera on a common viewport, without prior resampling. Having different images with different native pixel grids sharing the same map would require the web-client to perform real-time reprojection. Client-side reprojection should be possible e.g., with version 3 of the {\sc OpenLayers} library\footnote{\url{http://ol3js.org/}}}.
\end{itemize}

\section{Acknowledgments}
The authors would like to thank the anonymous referees whose comments helped not only in improving the clarity of this paper, but also the performance of the code.

CM wishes to acknowledge Prof. Joe Mohr for hospitality at USM, Munich, and the SkyMapper team, in particular Prof. Brian Schmidt, Dr. Patrick Tisserand and Dr. Richard Scalzo for support during her stay at MSO-ANU, Canberra where part of this work was completed. EB thanks Raphael Gavazzi, Val\'erie de Lapparent, and the Origin and Evolution of Galaxies group at IAP, Paris for financial support with the {\sc VisiOmatic} hardware, and Dr. Herv\'e Bouy at CAB, Madrid for providing content for the {\sc VisiOmatic} demos.

The {\sc VisiOmatic} client implements services provided by the Sesame Name Resolver and the VizieR catalog access tool developed at CDS, Strasbourg, France.

Some of our demonstration data are based on SDSS-III\footnote{\url{http://www.sdss3.org}} images. Funding for SDSS-III has been provided by the Alfred P. Sloan Foundation, the Participating Institutions, the National Science Foundation, and the U.S. Department of Energy Office of Science. SDSS-III is managed by the Astrophysical Research Consortium for the Participating Institutions of the SDSS-III Collaboration including the University of Arizona, the Brazilian Participation Group, Brookhaven National Laboratory, Carnegie Mellon University, University of Florida, the French Participation Group, the German Participation Group, Harvard University, the Instituto de Astrofisica de Canarias, the Michigan State/Notre Dame/JINA Participation Group, Johns Hopkins University, Lawrence Berkeley National Laboratory, Max Planck Institute for Astrophysics, Max Planck Institute for Extraterrestrial Physics, New Mexico State University, New York University, Ohio State University, Pennsylvania State University, University of Portsmouth, Princeton University, the Spanish Participation Group, University of Tokyo, University of Utah, Vanderbilt University, University of Virginia, University of Washington, and Yale University.

\bibliographystyle{model2-names}
\bibliography{invisu_article}

\end{document}